\documentclass[12pt]{article}

\ifx\pdfoutput\undefined
\usepackage[dvips,bookmarks]{hyperref}
\else
\usepackage{hyperref}
\fi
\hypersetup{colorlinks=false,bookmarksopen,bookmarksnumbered,citecolor=blue,
   pdfstartview=FitH}

\usepackage[pdftex]{graphicx}	
\usepackage{latexsym}

\usepackage{amssymb,amsfonts,amsmath}
\usepackage{graphicx} 
\usepackage{indentfirst}

 \usepackage{bbm}

\parskip=\medskipamount

\newcommand {\cD}{{\cal D}}

\newcommand {\cK}{{\cal K}}

\newcommand {\cM}{{\cal M}}
\newcommand {\cN}{{\cal N}}

\newcommand {\cR}{{\cal R}}

\newcommand {\cV}{{\cal V}}

\def\a{\alpha}

\def\b{\beta}

\def\d{\delta}
\def\e{\epsilon}

\def\g{\gamma}
\def\G{\Gamma}

\def\j{\psi}

\def\m{\mu}

\def\o{\omega}

\def\q{\theta}

\def\s{\sigma}

\def\x{\xi}
\def\z{\zeta}

\def\F{\Phi}
\def\J{\Psi}

\def\O{\Omega}

\def\S{\Sigma}

\def\ri{{\rm i}}
\def\re{{\rm e}}

\newcommand{\ad}{{\dot{\alpha}}}                           
\newcommand{\bd}{{\dot{\beta}}}                            
\newcommand{\ve}{\varepsilon}                            
\newcommand{\cDB}{{\bar\cD}}                            

\newcommand{\pa}{\partial}                           
\newcommand{\hf}{\frac12}

\newcommand{\be}{\begin{equation}}
\newcommand{\ee}{\end{equation}}
\newcommand{\bea}{\begin{eqnarray}}
\newcommand{\eea}{\end{eqnarray}}
\newcommand{\non}{\nonumber}
\newcommand{\bm}[1]{\mbox{\boldmath$#1$}}

\def\double #1{#1{\hbox{\kern-2pt $#1$}}}

\newcommand{\gd}{{\dot\g}}
\newcommand{\dd}{{\dot\d}}

\newif\ifdtup

\newcommand{\bsubeq}{\begin{subequations}}
\newcommand{\esubeq}{\end{subequations}}

\numberwithin{equation}{section}

\begin{document}
\begin{titlepage}
\begin{flushright}
December, 2012\\
\end{flushright}
\vspace{5mm}

\begin{center}
{\Large \bf 
Symmetries of curved superspace
}\\ 
\end{center}

\begin{center}

{\bf
Sergei M. Kuzenko
} \\
\vspace{5mm}

\footnotesize{
{\it School of Physics M013, The University of Western Australia\\
35 Stirling Highway, Crawley W.A. 6009, Australia}}  
\vspace{2mm}

\end{center}

\begin{abstract}
\baselineskip=14pt
The formalism to determine (conformal) isometries of a given curved superspace was elaborated 
almost two decades ago in the context of the old minimal formulation for $\cN=1$ supergravity in four dimensions (4D). This formalism is  universal, for it may readily be generalized 
to supersymmetric backgrounds associated with any supergravity theory formulated in superspace.
In particular, it has already been used to construct rigid supersymmetric field theories
in 5D $\cN=1$, 4D $\cN=2$ and 3D $(p,q)$ anti-de Sitter superspaces.  
In the last two years, there have appeared a number of publications devoted to the construction 
of supersymmetric 
backgrounds in off-shell 4D $\cN=1$ supergravity theories using component field considerations. 
Here we demonstrate how to read off the key results of these recent publications 
from the more general superspace approach developed in the 1990s.
We also present a universal superspace setting  
to construct supersymmetric backgrounds, 
which is applicable to any of the known off-shell formulations for $\cN=1$ supergravity.  
This approach is based on the realizations of the new minimal and non-minimal  supergravity theories 
as super-Weyl invariant couplings of the old minimal supergravity to certain conformal compensators. 
\end{abstract}

\vfill

\vfill
\end{titlepage}

\newpage
\renewcommand{\thefootnote}{\arabic{footnote}}
\setcounter{footnote}{0}

%


\section{Introduction}
\setcounter{equation}{0}

With the motivation to elaborate supersymmetric quantum field theory in curved space, 
almost two decades ago a formalism was  developed  \cite{BK}
to determine (conformal) isometries of a given curved superspace originating within the old minimal formulation for $\cN=1$ supergravity in four dimensions (4D).\footnote{Historically, 
this supergravity formulation was first constructed by Wess and Zumino in superspace \cite{WZ} (see also \cite{GWZ}), and soon after it was independently developed 
using the component tensor calculus \cite{old}. The superspace \cite{WZ} and the component 
\cite{old} approaches to  old minimal supergravity are equivalent, for the latter
can readily be deduced from the former  \cite{Zumino,WZ2,RL,GGRS} (see \cite{WB} for a review).}
As simple illustrations  of
the formalism,  it was used in \cite{BK} to compute (i) the conformal Killing supervector fields
of any conformally flat $\cN=1$ superspace; and (ii) the Killing supervector fields of
$\cN=1$ AdS superspace. The approach presented in \cite{BK} is  universal, for  
in principle it may  be generalized to supersymmetric backgrounds associated with any supergravity theory formulated in superspace.
In particular, it has already been used to construct rigid supersymmetric field theories
in 5D $\cN=1$ \cite{KT-M07}, 4D $\cN=2$ \cite{KT-M08,BKLT-M}
and 3D $(p,q)$ anti-de Sitter  \cite{KT-M11,KLT-M12,BKT-M}  superspaces. 

Recently, numerous publications have appeared devoted to the construction 
of supersymmetric backgrounds associated with  the old minimal and the new minimal supergravities 
using component field considerations 
\cite{FS,Jia:2011hw,Samtleben:2012gy,Klare:2012gn,DFS,KMTZ,Liu:2012bi,Dumitrescu:2012at,Kehagias:2012fh}.
These backgrounds are simply curved (pseudo) Riemannian spaces that allow unbroken rigid supersymmetries. The techniques used in these publications make no use of the superspace 
formalism of \cite{BK}.
However, since the rigid supersymmetry transformations are special isometry transformations 
of a given curved superspace, there should exist a simple procedure to derive the key 
component results of 
\cite{FS,Jia:2011hw,Samtleben:2012gy,Klare:2012gn,DFS,KMTZ,Liu:2012bi,Dumitrescu:2012at,Kehagias:2012fh}
from the more general superspace construction of \cite{BK}, for the latter allows one 
to determine all the isometries.  
One of the goals of the present note is to work out such a procedure. 
Our second, more important goal is to present a universal superspace setting, which can be used for  
any of the known off-shell formulations for $\cN=1$ supergravity, 
to determine the isometries of curved backgrounds.
This novel approach can immediately be generalized to all known off-shell supergravities 
in diverse dimensions, including the important cases of 3D $\cN\leq 4$, 
4D $\cN=2$ and 5D $\cN=1$ supergravity theories.   

It should be mentioned that a considerable amount of the results in 
\cite{FS,Jia:2011hw,Samtleben:2012gy,Klare:2012gn,DFS,Liu:2012bi,Dumitrescu:2012at,Kehagias:2012fh}
are devoted to supersymmetric backgrounds with Euclidean signature. 
Our analysis is restricted to curved space-times allowing unbroken supersymmetry. 

This paper is organized as follows. In section 2 we briefly review the superspace geometry of 
old minimal supergravity following the notation and conventions adopted in \cite{BK}.
Section 3 contains a summary of the main properties of the (conformal) Killing supervector
fields of a curved superspace derived in \cite{BK}. 
In section 4 we study those supergravity backgrounds without covariant fermionic fields
which allow some unbroken (conformal) supersymmetries. 
Section 5 describes a universal superspace setting 
to construct supersymmetric backgrounds, which is applicable to 
any of the known off-shell formulations for $\cN=1$ supergravity. 
Concluding comments are given in section 6. 

\section{The Wess-Zumino superspace geometry}

In describing the Wess-Zumino superspace geometry (see \cite{WB} for a review), 
we follow the notation and conventions of \cite{BK}.\footnote{These conventions 
are nearly identical to those
of Wess and Bagger \cite{WB}. To convert the notation of \cite{BK} to that of \cite{WB}, one
replaces $R \rightarrow 2 R$, $G_{\alpha \ad} \rightarrow 2 G_{\alpha \ad}$, and
$W_{\alpha \beta \gamma} \rightarrow 2 W_{\alpha \beta \gamma}$.
In addition, the vector derivative has to be changed by the rule 
$\cD_a \to \cD_a +\frac{1}{4}\ve_{abcd}G^b M^{cd}$, where $G_a$ corresponds to \cite{BK}. 
Finally, the spinor Lorentz generators $(\s_{ab})_\a{}^\b$ and 
$({\tilde \s}_{ab})^\ad{}_\bd$  used in \cite{BK} have an extra minus sign as compared with \cite{WB}, 
specifically $\s_{ab} = -\frac{1}{4} (\s_a \tilde{\s}_b - \s_b \tilde{\s}_a)$ and 
 $\tilde{\s}_{ab} = -\frac{1}{4} (\tilde{\s}_a {\s}_b - \tilde{\s}_b {\s}_a)$.  } 
In particular,  the coordinates of $\cN=1$ curved superspace $\cM$ are denoted 
$z^M = (x^m, \q^\m , {\bar \q}_{\dot \m})$.
The superspace geometry is described 
by covariant derivatives of the form
\bea
\cD_A &=& (\cD_a , \cD_\a ,\cDB^\ad ) = E_A + \O_A~.
\eea
Here $E_A$ denotes the inverse vielbein, 
$E_A = E_A{}^M  \pa_M $,
and $\O_A$  the Lorentz connection, 
\bea
\O_A = \hf\,\O_A{}^{bc}  M_{bc}
= \O_A{}^{\b \g} M_{\b \g}
+\O_A{}^{\bd \gd} {\bar M}_{\bd \gd} ~,
\eea
with $M_{bc} \Leftrightarrow ( M_{\b\g}, {\bar M}_{\bd \gd})$
the Lorentz generators.
The covariant derivatives obey the following anti-commutation relations:
\begin{subequations}\label{algebra}
\bea
&& {} \qquad \{ \cD_\a , {\bar \cD}_\ad \} = -2{\rm i} \cD_{\a \ad} ~, 
\non \\
\{\cD_\a, \cD_\b \} &=& -4{\bar R} M_{\a \b}~, \qquad
\{ {\bar \cD}_\ad, {\bar \cD}_\bd \} =  4R {\bar M}_{\ad \bd}~,  \\
\left[ { \bar \cD}_{\ad} , \cD_{ \b \bd } \right]
     & = & -{\rm i}{\ve}_{\ad \bd}
\Big(R\,\cD_\b + G_\b{}^{\dot{\g}}  \cDB_{\dot{\g}}
-(\cDB^\gd G_\b{}^{\dot{\d}})
{\bar M}_{\gd \dot{\d}}
+2W_\b{}^{\g \d}
M_{\g \d} \Big)
- {\rm i} (\cD_\b R)  {\bar M}_{\ad \bd}~,~~~~~~~  \\
\left[ \cD_{\a} , \cD_{ \b \bd } \right]
     & = &
     {\rm i}
     {\ve}_{\a \b}
\Big({\bar R}\,\cDB_\bd + G^\g{}_\bd \cD_\g
- (\cD^\g G^\d{}_\bd)  M_{\g \d}
+2{\bar W}_\bd{}^{\gd \dot{\d}}
{\bar M}_{\gd \dot{\d} }  \Big)
+ {\rm i} (\cDB_\bd {\bar R})  M_{\a \b}~,  \\
\big[ \cD_{\a\ad} , \cD_{\b \bd} \big] &=& \ve_{\ad \bd} \j_{\a \b} +\ve_{\a\b} \j_{\ad\bd} ~,
\label{algebra_d}
\eea
where
\bea
\j_{\a\b} &:=& -\ri G_{( \a }{}^\gd \cD_{\b)\gd} +\hf (\cD_{(\a} R) \cD_{\b )} 
+\hf (\cD_{(\a }G_{\b )}{}^\gd )\bar \cD_\gd +W_{\a\b}{}^\g \cD_\g \non \\
&& +\frac{1}{4} \big((\bar \cD^2 -8R) \bar R \big)M_{\a\b} + (\cD_{( \a} W_{\b )}{}^{\g\d } )M_{\g\d}
-\hf (\cD_{( \a }\bar \cD^\gd G_{\b )}{}^\dd)  \bar M_{\gd \dd}~,  \label{algebra_e}\\
\j_{\ad\bd} &:=&- \ri G_{\g (\ad } \cD^\g {}_{\bd )} -\hf (\bar \cD_{(\ad} \bar R) \bar \cD_{\bd )}
-\hf (\bar \cD_{(\ad } G^\g{}_{\bd )} )\cD_\g -\bar W_{\ad \bd }{}^\gd \bar \cD_\gd \non \\
&& +\frac{1}{4}\big( (\cD^2 - 8 \bar R)  R\big) \bar M_{\ad \bd}
-(\bar \cD_{(\ad} \bar W_{\bd)}{}^{\gd \dd} )\bar M_{\gd \dd} 
+ \hf (\bar \cD_{(\ad} \cD^\g G^\d{}_{\bd)} )M_{\g\d}~. 
\label{algebra_f}
\eea 
\end{subequations}
The torsion tensors $R$, $G_a = {\bar G}_a$ and
$W_{\a \b \g} = W_{(\a \b\g)}$ satisfy the Bianchi identities
\begin{subequations}
\bea
\cDB_\ad R&=& 0~, \qquad \cDB_\ad W_{\a \b \g} = 0~, \label{2.4a} \\
\cDB^\gd G_{\a \gd} &=& \cD_\a R~, \label{2.4b} \\
\cD^\g W_{\a \b \g} &=& {\rm i} \,\cD_{(\a }{}^\gd G_{\b) \gd}~.
\label{2.4c} 
\eea
\end{subequations}

A supergravity gauge transformation is defined to act on the covariant derivatives
and any tensor superfield $U$ (with its indices suppressed) by the rule
\begin{subequations}
\bea
\d_\cK \cD_A = [\cK, \cD_A] ~, \qquad \d_\cK U = \cK U~.
\eea
Here the gauge parameter $\cK$ has the explicit form 
\bea
\cK = \x^B \cD_B + K^{\g\d} M_{\g \d} + \bar K^{  \gd \dd} \bar M_{ \gd \dd} = \bar \cK
\label{K}
\eea
\end{subequations}
and describes a general coordinate transformation generated 
by the supervector field $\x^B$ as well as a local Lorentz transformation generated by 
the symmetric spinor $K^{ \g \d} + \x^B \O_B{}^{\g \d} $ and its conjugate. 

The algebra of covariant derivatives \eqref{algebra} 
is invariant under super-Weyl transformations \cite{HT} 
\begin{subequations} 
\label{superweyl}
\bea
\d_\s \cD_\a &=& ( \hf \s - {\bar \s} )  \cD_\a - (\cD^\b \s) \, M_{\a \b}  ~, \\
\d_\s \bar \cD_\ad & = & ( \hf {\bar \s} - \s  )
\bar \cD_\ad -  ( \bar \cD^\bd  {\bar \s} )  {\bar M}_{\ad \bd} ~,\\
\d_\s \cD_{\a\ad} &=& -\hf( \s +\bar \s) \cD_{\a\ad} 
-\frac{\ri}{2} (\bar \cD_\ad \bar \s) \cD_\a - \frac{\ri}{2} ( \cD_\a  \s) \bar \cD_\ad \non \\
&& - (\cD^\b{}_\ad \s) M_{\a\b} - (\cD_\a{}^\bd \bar \s) \bar M_{\ad \bd}~,
\eea
\end{subequations}
with the scalar parameter $\s$ being covariantly chiral, 
\bea
\bar \cD_\ad \s =0~,
\label{sigma_chiral}
\eea
provided the torsion tensors transform\footnote{The super-Weyl transformation of $G_{\a\ad}$
given in \cite{BK}, eq. (5.5.14), contains a typo.}
 as follows:
\begin{subequations} 
\bea
\d_\s R &=& -2\s R -\frac{1}{4} (\bar \cD^2 -4R ) \bar \s ~, \\
\d_\s G_{\a\ad} &=& -\hf (\s +\bar \s) G_{\a\ad} +\ri \cD_{\a\ad} (\bar \s- \s) ~, \label{s-WeylG}\\
\d_\s W_{\a\b\g} &=&-\frac{3}{2} \s W_{\a\b\g}~.
\eea
\end{subequations}

Let  ${\frak D}_{A}=({\frak D}_a,{\frak D}_\a,\bar {\frak D}^\ad)$ be another set of superspace covariant derivatives  
which describe a curved supergravity background.
The two superspace geometries, which are associated with $\cD_A$ and ${\frak D}_A$, 
are said to be conformally related  if their covariant derivatives  are related by a 
finite super-Weyl transformation
\bsubeq \label{2.9}
\bea
{\frak D}_\a&=&\re^{\hf \o-\bar{\o}}\Big(\cD_\a-(\cD^\b\o)M_{\a\b}\Big)
~,
\\
\bar {\frak D}_\ad&=&\re^{\hf \bar \o-\o}\Big(\bar \cD_\ad
-(\bar \cD^\bd\bar \o)\bar{M}_{\ad\bd}\Big)
~,
\\
{\frak D}_{\a\ad}&=&\frac{\ri}{2}\{ {\frak D}_\a,\bar {\frak D} _\bd\}~,
\eea
\esubeq
where $\o$ is a covariantly chiral scalar, $\bar \cD_\ad \o =0$.

\section{(Conformal) Killing supervector fields}

Let us fix a  curved background superspace $\cM$. 
In accordance with \cite{BK}, 
a supervector field $\x= \x^B E_B$ on $\cM$
is called conformal Killing if there exists a 
symmetric spinor $K^{\g\d}$ and a covariantly chiral scalar $\s$ such that 
\bea
(\d_\cK + \d_\s) \cD_A =0~.
\label{conf_Killing}
\eea
In other words, the  coordinate transformation generated by $\x$ can be accompanied by certain Lorentz
and super-Weyl transformations such that the superspace geometry does not change.

As demonstrated in \cite{BK},  all information about the conformal Killing supervector field 
is encoded in the special case of  eq. \eqref{conf_Killing}
with $A=\a$. Making use of the variation
\bea
\d_\cK \cD_\a&=&\Big({K_\a}^\b-\cD_\a \xi^\b
-  \frac{ \ri }{2} \xi_{\a  \bd}G^{\b \bd}\Big)\cD_\b
+  \Big(\cD_\a \bar\xi^{ \bd}+
  \frac{\ri}{2}{\xi_\a}^{\bd} \bar R\Big)\bar \cD_{\bd} \nonumber\\
&& +2\ri \Big({\d_\a}^\b \bar\xi^{ \bd}
-  \frac{\ri}{4} \cD_\a \xi^{\b  \bd}\Big)\cD_{\b  \bd} \nonumber\\
&& -\Big(\cD_\a K^{\b \g} + 4 {\d_\a}^{(\b}\xi^{\g)}\bar R
-  \frac{\ri}{2}{\d_\a}^{(\b}\xi^{\g) \,\gd }\bar \cD_{\gd} \bar R
-  \frac{\ri}{2} \xi_{\a  \ad} \cD^{(\b}G^{\g)\dot \a}\Big)M_{\b\g} \nonumber\\
&& -(\cD_\a \bar K^{\dot \b \dot \g } + \ri \xi_{\a \dot \a}
  \bar W^{\dot\a \dot\b \dot\g})\bar M_{\dot\b\dot\g}~, 
\eea
in conjunction with the super-Weyl transformation \eqref{superweyl}, 
we obtain a number of conditions on the parameters 
which  can be split in two groups. The first group consists of the  following equations 
\begin{subequations} \label{3.3}
\bea
\d_\a{}^\b \bar \x^\bd &=& \frac{\ri}{4} \cD_\a\x^{\b\bd} 
\quad \Rightarrow \quad {\bar \x}^\ad = \frac{ \ri }{8} \cD_\a \x^{\a \ad}
~, \\
K_{\a\b} &=& \cD_{(\a } \x_{\b)}  -\frac{\ri}{2} \x_{ (\a}{}^\bd G_{\b) \bd}~, \\
\s &=& \frac{1}{3} (\cD^\a \x_\a +2 \bar \cD_\ad \bar \x^\ad - {\ri}  \x^{a} G_{a} )~,
\label{3.3c}
\eea
\end{subequations}
and their conjugates. Eq.  \eqref{3.3c} has to be taken in conjunction  
with the chirality condition,  eq. \eqref{sigma_chiral},
obeyed by the super-Weyl parameter.  
The meaning of the relations \eqref{3.3} is that the parameters $\x^\a$, $K^{\a \b}$ and $\s$ 
are completely determined in terms of the real vector $\x^a$ and its covariant derivatives.  
This is why  we may also use the notation $\cK = \cK[\x]$, 
and similarly for the Lorentz and super-Weyl parameters, e.g. $\s = \s[\x]$.

The second group comprises the following equations and their conjugates:
\begin{subequations} \label{3.4}
\bea
\cD_\a \bar \x_\ad &=& - \frac{\ri}{2} \x_{\a\ad} \bar R~, \\
\bar \cD_\ad K^{\b\g} &=& \ri \x_{\a\ad} W^{\a\b\g}~,\\
\cD_\a K^{\b\g} &=& - \d_\a{}^{(\b} \cD^{\g)} \s - 4 \d_\a{}^{(\b} \x^{\g)} \bar R 
+\frac{ \ri }{2} \d_\a{}^{ (\b} \x^{\g )\gd} \bar \cD_\gd \bar R +\frac{\ri}{2} \x_{\a\ad} \cD^{(\b} G^{\g) \ad}~.
\eea
\end{subequations}
These relations allow us to express multiple covariant spinor derivatives of the parameters 
in terms of the parameters.\footnote{In the non-conformal case, which corresponds to $\s=0$, 
the first spinor covariant derivatives of the parameters $\x^B$, $K^{\b \g}$ and $\bar K^{\bd \gd}$ 
are linear combinations of these parameters.}

Since the real vector $\x^a$ is the only independent parameter, there should exist a 
closed-form equation obeyed by $\x^a$. It has the form
\bea
\cD_{(\a} \x_{\b) \bd} =0~.
\label{master}
\eea
Simple corollaries of this equation\footnote{The equation \eqref{master} is analogous 
to the conformal Killing equation, $\nabla_{(\a}{}^{( \ad} V_{\b )}{}^{\bd )} =0$,
on a (pseudo) Riemannian four-dimensional manifold.}
include the linearity condition 
\bea
(\cD^2 +2 \bar R) \x_a = 0 ~,
\eea
and the conformal Killing equation
\bea
\cD_a \x_b +\cD_b \x_a = \hf \eta_{ab} \cD^c \x_c~.
\eea
As shown in \cite{BK}, all information about the conformal Killing supervector field is encoded 
in the master equation \eqref{master}. Specifically, if this equation holds and  the definitions \eqref{3.3}
are adopted, then the consistency conditions \eqref{3.4} are identically satisfied, 
and the super-Weyl parameter $\s[\x] $ is covariantly chiral. 
As a result,  an alternative definition of the conformal Killing supervector field can be given.
It is a real supervector field 
\bea
\x= \x^A E_A ~, \qquad \x^A = \Big( \x^a, -\frac{\ri }{8} \bar \cD_\b \x^{\a \bd} , 
-\frac{\ri}{8} \cD^\b \x_{\b \ad} \Big)
\eea
which obeys the master equation \eqref{master}.

If $\x_1$ and $\x_2$ are two conformal Killing supervector fields, their 
Lie bracket $[\x_1, \x_2]$ is a conformal Killing supervector field \cite{BK}. 
It is obvious that, for any real $c$-numbers $r_1$ and $r_2$, the linear combination
 $r_1 \x_1 + r_2 \x_2$ is a  conformal Killing supervector field. 
Thus the set of all conformal Killing supervector
fields is a super Lie algebra. The conformal Killing supervector fields generate symmetries 
of a super-Weyl invariant field theory on $\cM$. 

We need to recall one more result from \cite{BK}. Suppose we have another curved superspace
$\frak M$ that is conformally related to $\cM$. This means that the covariant derivatives
 $\cD_A$ and ${\frak D}_A$, which correspond to $\cM$ and $\frak M$ respectively, 
 are related to each other according to the rule \eqref{2.9}. It turns out that  the two superspaces 
$\cM$ and $\frak M$ have the same conformal Killing supervector fields. Given such a 
supervector field $\x$, it can be represented in two different forms
\bea
\x= \x^A   E_A = {\bm \x}^A {\frak E}_A~,
\eea
where ${\frak E}_A$ is the inverse vielbein associated with the covariant derivatives 
${\frak D}_A$. Then the super-Weyl parameter $\s[\x]$ and $\s [{\bm \x}]$ are related 
to each other as follows
\bea
\s[{\bm \x}] = \s[\x] - \x\, \o~.
\eea
The derivation of this result is given in \cite{BK}.

A Killing supervector field $\x$ on $\cM$
is a conformal Killing supervector field with the additional property
$\s[\x]=0$, or equivalently 
\bea
\d_\cK \cD_A = [\cK , \cD_A]=0~.
\label{Killing}
\eea
The condition that  the super-Weyl parameter \eqref{3.3c} be equal to zero  is 
\bea
\cD^\a \x_\a = -\ri G^a \x_a~.
\eea
If $\x_1$ and $\x_2$ are Killing supervector fields, their Lie bracket $[\x_1, \x_2]$ 
is a  Killing supervector field. Thus the set of all  Killing supervector
fields forms a super Lie algebra. The Killing supervector fields generate the isometries of $\cM$,  
and symmetries of a field theory on $\cM$. 

To study supersymmetry transformations at the component level, 
it is useful to spell out one of the implications of eq. \eqref{superweyl}
with $A= a$. Specifically, we consider the equation $(\d_\cK + \d_\s) \cD_{\a\ad} =0$ 
and read off its part proportional to a linear combination of the spinor covariant derivatives $\cD_\b$. 
The results is
\bea
0=\cD_{\a\ad} \x_\b &-& \frac{\ri}{2}  \ve_{\a \b} \bar \cD_\ad \bar \s
+\ri \x_\a G_{\b \ad} - \ri  \ve_{\a\b}\bar \x_\ad R \non \\
&-& \frac{1}{4} \x_{\b\ad} \cD_\a R - \hf \x^\g{}_\ad W_{\a\b\g}
+\frac{1}{4}\x_\a{}^\gd \bar \cD_\ad G_{\b \gd}~.
\label{3.8}
\eea
In the case of isometry transformations on $\cM$, we have to set $\s=0$. 
Eq. \eqref{3.8} will play a fundamental role in our subsequent analysis.

\section{Supersymmetric backgrounds} \label{section4}

We wish to look for those curved backgrounds which admit some  unbroken 
(conformal) supersymmetries. By definition, such a superspace possesses 
a (conformal) Killing supervector field $\x^A$  with the property
\bea
\x^a|=0 ~, \qquad \e^\a := \x^\a | \neq 0~, 
\eea
where $U|$ denotes the $\q,\bar \q$ independent part
of a tensor superfield $U(z)= U(x^m,\q^\m, \bar \q_{\dot \m}) $, 
\bea
U|:=  U|_{\q^\m=\bar \q_{\dot \m}=0} ~.
\eea
We will refer to the field $U|$ as the bar-projection of $U$.\footnote{In the case of $\cN=2$ supergravity formulated in 
superspace, it is also useful to consider a partial superspace reduction $\cN=2 \to \cN=1$
\cite{Gates:1983ie}.} 
Our analysis will be restricted to supergravity  backgrounds without covariant fermionic fields, 
that is
\bea
\cD_\a R |=0~, \qquad \cD_\a G_{\b\bd}|=0~, \qquad W_{\a \b \g}|=0~.
\label{4.3}
\eea
This means that the gravitino can completely be gauged away such that the bar-projection of the 
vector covariant derivative\footnote{The bar-projection of a covariant derivative, $\cD_A|$, 
is defined by the rule $ (\cD_A | U)|:= (\cD_A U)|$, for any tensor superfield $U$.
The bar-projection of a product of several covariant derivatives, $\cD_{A_1} \cdots \cD_{A_n}|$, 
is defined similarly.}
is
\bea
\widetilde{\nabla}_a:=\cD_a| =  \nabla_a +\frac{1}{6} \ve_{abcd} b^b M^{cd}~, \qquad
\nabla_a = e_a{}^m \pa_m +  \hf \o_{a}{}^{cd}M_{cd} ~,
\label{4.4} 
\eea
where $\nabla_a$ denotes the ordinary torsion-free covariant derivative, 
\bea
[ \nabla_a, \nabla_b] = \hf \cR_{ab}{}^{cd}M_{cd} ~, 
\eea
and the vector field $b_a $ is one of the auxiliary fields $M$, $\bar M$ and $b_a$, 
which correspond to the old minimal supergravity and are defined as\footnote{To simplify comparison
with the results of \cite{FS}, here we make use of the same definition of the auxiliary fields 
as in \cite{FS},  following \cite{WB}.
These are related to the supergravity auxiliary fields $\mathbb B$ and ${\mathbb A}_a$
used \cite{BK} by the rule: $M=- \bar {\mathbb B}$ and $b_a = - 2{\mathbb A}_a$.}
\bea
R | = -\frac{1}{3} M~, \qquad G_a|= -\frac{2}{3} b_a~.
\eea
 
\subsection{Conformal supersymmetry}
 
Let us first determine the conditions for unbroken conformal supersymmetry. For this, 
we consider the $\q,\bar \q$  independent part of the equation \eqref{3.8}.
With the definition 
\bea
\cD_\a \s | = -\frac{2}{3} \z_\a~, 
\eea
the result is
\bea
2 \nabla_a \e_\b 
-\frac{\ri }{3} \Big\{ (\s_a \bar \z)_{\b} +  (\s_a \bar \e)_{\b}  M
-2(\s_{ac} \e)_\b b^c +2 b_a \e_\b \Big\} =0~.
\label{conformal_spinor}
\eea 
This equation allows one to express the conformal spinor parameter $\bar \z_\ad$ in terms of $\e_\a$
and its conjugate.

Eq. \eqref{conformal_spinor} can be rewritten
in a different and more illuminating form if we 
introduce the first-order operator
\bea
{\mathfrak D}_a \e_\b := (\nabla_a -\frac{\ri}{2} b_a ) \e_\b
\eea
which can be viewed as a  U(1) gauge covariant derivative.
Then one may see that \eqref{conformal_spinor} is equivalent to 
\bea
2{\mathfrak D}_a \e_\b - \frac{\ri }{3} \Big\{ (\s_a \bar \z)_{\b} +  (\s_a \bar \e)_{\b}  M
+ (\s_a \tilde{\s}_c \e)_\b b^c \Big\} =0  ~.
\label{4.11}
\eea
Expressing here  $\bar \z_\ad$ in terms of $\e_\a$
and its conjugate leads to the equation
\bea
{\mathfrak D}_a \e_\b +\frac{1}{4} (\s_a\tilde{\s}^c {\mathfrak D}_c \e)_\b =0~,
\eea
which is equivalent to
\bea
{\mathfrak D}_{\a\gd } \e_\b + {\mathfrak D}_{\b \gd } \e_\a =0~.
\label{conformal_spinor2} 
\eea
We conclude that  $\e_\b$ is a  charged conformal Killing spinor. 
Given a non-zero solution $\e_\a (x)$ of \eqref{conformal_spinor2}, 
as well as a non-zero complex number
$z \in {\mathbb C} $,  it is obvious that $z \,\e_\a (x)$ is also a solution of the same equation. 
We conclude that the minimal amount of conformal supersymmetry is two supercharges, 
which agrees with the conclusions in \cite{KMTZ}.  

If $\e_\b$ is a commuting  conformal Killing spinor obeying the equation  
\eqref{conformal_spinor2},  the {\it null} vector 
${\cV}_{\b\bd} := \e_\b \bar \e_\bd$ is a conformal Killing vector field, 
\bea
\nabla_{(\a}{}^{( \ad} \cV_{\b )}{}^{\bd )} =0~.
\eea
Thus we have re-derived one of the key  results of \cite{KMTZ}.\footnote{It was 
demonstrated in \cite{KMTZ} that
$\cM$ possesses a conformal Killing spinor if and only if it has a null conformal Killing vector.}

\subsection{Rigid supersymmetry} 

In the non-conformal case, 
setting $\bar \z^\ad=0$ in \eqref{conformal_spinor}
gives  the equation for unbroken rigid supersymmetry
\bea
2 \nabla_a \e_\b 
-\frac{\ri }{3} \Big\{ (\s_a \bar \e)_{\b}  M
-2(\s_{ac} \e)_\b b^c +2 b_a \e_\b \Big\} =0~.
\label{rigid_spinor}
\eea 
This equation coincides with that given in \cite{FS} keeping in mind the fact that 
the matrices $\s_{ab}$ used in \cite{FS} differ in sign from ours. 

\subsection{Curved spacetimes admitting four supercharges}

We now turn to deriving those conditions on the background geometry which guarantee that  
the spacetime under consideration possesses nontrivial solutions of eq. \eqref{rigid_spinor}
giving rise to exactly four supercharges.  
The main idea of our analysis below is that the conditions \eqref{4.3} must be supersymmetric. 

To start with, consider the identity
\bea 
0&=& \d_\cK \cD_\a R = \x^D \cD_D \cD_\a + K_\a{}^\g \cD_\g R \non \\
&=& \x^c \cD_c \cD_\a R -\hf \x_\a \cD^2 R +2\ri \bar \x^\gd \cD_{\a\gd} R  +K_\a{}^\g \cD_\g R~.
\eea
The bar-projection of this relation is 
\bea
\e_\a \cD^2 R| - 4\ri \bar \e^\gd \nabla_{\a\gd} R| =0~.
\eea
This is equivalent to 
\begin{subequations}
\bea
\cD^2 R|&=& 0~,  \label{D_R1} \\
\nabla_{a} M &=& 0~. \label{D_R2}
\eea
\end{subequations}
The complete expression for  $\cD^2 R|$ in terms of the supergravity fields 
can be found in, e.g., \cite{BK} and \cite{WB}. We will not need this expression, 
for the condition \eqref{D_R1} proves to follow from a more general result to be derived shortly. 
Eq. \eqref{D_R2} means that $M$ is a constant parameter. 

The next condition to analyze is
\bea
0&=& \d_\cK \cD_\a G_{\b\bd} = \x^D \cD_D \cD_\a + K_\a{}^\d \cD_\d G_{\b\bd} 
+ K_\b{}^\d \cD_\a G_{\d\bd} +\bar K_\bd{}^\dd \cD_\a G_{\b\dd} ~.
\eea
This leads to 
\bea
\e^\d \cD_\d \cD_\a G_{\b \bd}| - \bar \e^{\dd} \bar \cD_\dd \cD_\a G_{\b\bd}| =0~,
\eea
and hence 
\begin{subequations}
\bea
\cD_\d \cD_\a  G_{\b\bd} |&=&\bar  \cD_\dd \bar \cD_\ad  G_{\b\bd} | = 0~,  \\
\bar \cD_\dd \cD_\a  G_{\b\bd} |&=&  \cD_\d \bar \cD_\ad  G_{\b\bd} |
=0~.
\eea
\end{subequations}
These conditions\footnote{The complete expression for 
$\bar \cD_{(\ad} \cD^{(\g } G^{\d)}{}_{\bd)} |$ is given in \cite{BK}. We do not need 
it for our analysis.}  
imply, in particular, that  $R| G_{\b\bd}  | =0 $ and $\cD_{\a \gd}G_{\b\bd}|=0$, 
or equivalently
\begin{subequations}  \label{4.14}
\bea
  M b_{c} &=&0~,  \label{4.14a}\\
\nabla_a b_c &=&  0~. \label{4.14b}
\eea
\end{subequations}
We conclude that the vector field $b_a$ is covariantly constant.  Eq. \eqref{4.14a}
holds if $M=0$ or $b_a =0$. 

The last condition to analyze is 
\bea
0 = \d_\cK W_{\a\b\g} = \x^D \cD_D W_{\a\b\g} + 3 K^\d{}_{(\a} W_{\b\g)\d} ~.
\eea
It leads to $\e^\d \cD_\d W_{\a\b\g} |=0$, and hence
\bea
\cD_\d W_{\a\b\g} |=0~. 
\label{4.16}
\eea
In virtue of the Bianchi identity \eqref{2.4c}, the  condition $\cD^\g W_{\a\b\g} |=0$ 
automatically holds if \eqref{4.14b} is satisfied. The nontrivial part of \eqref{4.16} is 
\bea
\cD_{(\d} W_{\a\b\g )} |=0~.
\label{4.24}
\eea
The complete expression for $\cD_{(\d} W_{\a\b\g )} |$ is given in \cite{BK}.  
Since the gravitino is absent, eq. \eqref{4.24}  
tells us that the Weyl tensor is equal to zero, 
\bea
C_{\a\b\g\d} =0~.
\label{4.25}
\eea
 As a result, the space-time is  conformally flat.

The above results can be used to read off the Riemann tensor. For this we compute the bar-projection
$[\cD_a, \cD_b]|$ in two different ways. First of all, we can make use of \eqref{4.4} to obtain 
\bea
[\cD_a, \cD_b]|= [\widetilde{\nabla}_a , \widetilde{\nabla}_b] ~.
\eea
The right-hand side has to be expressed in terms of 
the torsion-free covariant derivatives $\nabla_a$. Direct calculations give 
\bea
[\widetilde{\nabla}_a , \widetilde{\nabla}_b]  V_c&=& 
[{\nabla}_a , {\nabla}_b] V_c -\frac{2}{3} \ve_{ab d e} b^d \nabla^e V_c 
\non \\ 
&+&
\frac{1}{9} \Big\{ b_c (b_a \eta_{bd} - b_b \eta_{a d} ) 
- b_d (b_a \eta_{bc} - b_b \eta_{a c} ) 
-b^2 (\eta_{ac} \eta_{bd} - \eta_{ad} \eta_{bc} \Big\}V^d~.~~~~
\eea
On the other hand, we can evaluate $[\cD_a, \cD_b]|$ by making use of eqs. 
\eqref{algebra_d} -- \eqref{algebra_f}. This gives
\bea
[\cD_a, \cD_b]| V_c &=&  -\frac{2}{3} \ve_{ab d e} b^d \widetilde{\nabla}^e V_c 
-\frac{1}{9} M \bar M ( \eta_{ac}\eta_{bd} - \eta_{ad} \eta_{bc} )V^d \non \\
&=&  -\frac{2}{3} \ve_{ab d e} b^d {\nabla}^e V_c 
-\frac{1}{9} M \bar M ( \eta_{ac}\eta_{bd} - \eta_{ad} \eta_{bc} )V^d \non\\
&+&
\frac{2}{9} \Big\{ b_c (b_a \eta_{bd} - b_b \eta_{a d} ) 
- b_d (b_a \eta_{bc} - b_b \eta_{a c} ) 
-b^2 (\eta_{ac} \eta_{bd} - \eta_{ad} \eta_{bc} \Big\}V^d~.~~~~
\eea
We end up with  the Riemann curvature 
\bea
R_{abcd} = &&\frac{1}{9} \Big\{ b_c (b_a \eta_{bd} - b_b \eta_{ad}) 
-  b_d (b_a \eta_{bc} - b_b \eta_{ac})  -b^2 (\eta_{ac}\eta_{bd} - \eta_{ad} \eta_{bc} )\Big\} \non \\
&-&\frac{1}{9} M \bar M ( \eta_{ac}\eta_{bd} - \eta_{ad} \eta_{bc} )~.
\label{4.28}
\eea
The Ricci tensor is 
\bea
R_{ab} = \frac{2}{9} ( b_a b_b - b^2 \eta_{ab}) - \frac{1}{3} M \bar M \eta_{ab}~.
\label{4.29}
\eea

The conditions \eqref{D_R2}, \eqref{4.14a}, \eqref{4.14b} and \eqref{4.25} coincide with 
those given in \cite{FS} without derivation. 
Our expression for the Ricci tensor \eqref{4.29} differs from that  given in \cite{FS} by an overall sign.
This difference is due to the different definitions of the Riemann tensor used in \cite{FS} 
and in the present paper. 
The superspace techniques 
make the derivation of the conditions
 \eqref{D_R2}, \eqref{4.14a}, \eqref{4.14b}, \eqref{4.25} and \eqref{4.29} 
almost trivial.

Since $\cM$ is conformally flat, the corresponding algebra of conformal Killing supervector 
fields coincides with that of Minkowski superspace, ${\frak su}(2,2|1)$.

\section{Variant off-shell formulations for supergravity}\label{section5}

As is well known, there exist three off-shell formulations for $\cN=1$ supergravity in four dimensions:
(i) the old minimal formulation ($n=-1/3)$ developed first by Wess and Zumino 
using superspace techniques
\cite{WZ} and soon after in the component field approach \cite{old};\footnote{The linearized version of old minimal supergravity was constructed in  \cite{OS1,FZ}.}  
(ii) the new minimal formulation ($n=0)$
developed by Sohnius and West
\cite{new2};\footnote{The linearized version of new minimal \cite{new1}
supergravity  appeared much earlier than \cite{new2}.}
(iii) the non-minimal formulation  ($n\neq -1/3, 0$) pioneered by Breitenlohner  
\cite{non-min}, who used superspace techniques,  
and further developed to its modern form by Siegel and Gates  \cite{SG,GS80}. 
Breitenlohner's formulation \cite{non-min} is the oldest off-shell supergravity theory 
in four dimensions. 

Each off-shell formulation for $\cN=1$ supergravity can be realized as a super-Weyl invariant 
coupling of the old minimal supergravity ($n=-1/3$) to a 
{\it scalar} compensator $\J$ and its conjugate $\bar \J$ 
(if the compensator is complex) \cite{BK,FGKV,Kugo2,BKS} with a super-Weyl transformation of the form
\bea
\d_\s \J = - (p \s + q \bar \s) \J~, 
\eea
where $p$ and $q$ are fixed parameters which are determined by the off-shell structure of $\J$. 
The compensator is assumed to be nowhere vanishing.

In this super-Weyl invariant setting, the supergravity prepotentials include 
the compensator $\J$ and its conjugate. If we are interested in a fixed curved background, 
the supergravity gauge freedom and the super-Weyl invariance should be fixed in a convenient way to 
eliminate superfluous degrees of freedom. In particular, the super-Weyl invariance can be used to eliminate some of the component fields of $\J$ and its conjugate. 
The isometries of the resulting curved superspace are generated by those conformal Killing supervector
fields $\x= \x^B E_B  $, eq. \eqref{conf_Killing}, which leave the compensator invariant, that is 
\bea
\x^B \cD_B \J - (p \s + q \bar \s) \J =0 \quad \Longleftrightarrow \quad
(p \s + q \bar \s)  = \x^B \cD_B \ln \J ~.
\label{5.2}
\eea

\subsection{Old minimal supergravity}
The compensators in old minimal supergravity are a covariantly chiral scalar $\F$, 
$\bar \cD_\ad \F=0$, and its conjugate $\bar \F$. The super-Weyl transformation of $\F$ 
can conveniently be chosen to be 
\bea
\d_\s \F = - \s \F~,
\eea
and thus $p=1$ and $q=0$. 

The super-Weyl gauge freedom can be used to impose the gauge 
condition 
\bea
\F=1~.
\eea 
In this gauge the equation \eqref{5.2} becomes
\bea
\s =0~, 
\eea
and therefore the isometries are described by the Killing spinor equation \eqref{Killing}.

\subsection{New minimal supergravity}
In new minimal supergravity, the compensator $\frak L$ is a real covariantly linear scalar, 
\bea
(\bar \cD^2 -4R ){\frak L} =0 ~, \qquad \bar {\frak L} = {\frak L}~.
\eea
Such a superfield describes the $\cN=1$ tensor multiplet \cite{Siegel}.\footnote{The 
compensator for new minimal supergravity is often called the improved tensor multiplet \cite{deWR},
because the corresponding  action must be  super-Weyl invariant, which  corresponds 
to a uniquely determined self-coupling for $\frak L$, with the superfield Lagrangian being proportional 
to ${\frak L} \ln {\frak L}$.}
Its super-Weyl transformation is uniquely determined (see \cite{BK} for more details)
\bea
\d_\s {\frak L} = - (\s +\bar \s) {\frak L}~,
\eea
and thus $p=q=1$. 

In new minimal supergravity,  the super-Weyl gauge freedom 
may conveniently be fixed by imposing the conditions 
\begin{subequations} \label{5.8}
\bea
R| &=&0~, \label{5.8a}\\
{\frak L} |&=& 1~, \\
\cD_\a {\frak L}|&=&0~.
\eea
\end{subequations}
This leaves unbroken a U$(1)_R$ gauge symmetry generated by $\ri (\bar \s - \s)|$. The gauge field for this 
local symmetry is the auxiliary field $b_a$, in accordance with  eq. \eqref{s-WeylG}.

In the gauge \eqref{5.8}, 
there still remains a single component field contained in $\frak L$ that can be defined as
(see, e.g.,  \cite{BGG}):
\bea
-2 H_{\a \ad} := ([\cD_\a, \bar \cD_\ad] -2 G_{\a\ad}) {\frak L} \big|  
= [\cD_\a, \bar \cD_\ad] {\frak L}|  + \frac{4}{3} b_{\a\ad} ~. 
\eea
Here $H^a =\frac{1}{3!} \ve^{abcd}H_{bcd}$ is the Hodge-dual of the field strength 
of a gauge two-form.
Eq. \eqref{5.8a} means that $M=0$.
This auxiliary field is not present in new minimal supergravity.

The  Killing equation \eqref{5.2} corresponding to new minimal supergravity
has the form
\bea
(\s + \bar \s)  = \x^B \cD_B \ln {\frak L}~.
\label{5.7}
\eea 
We are in a position to derive an equation for unbroken rigid supersymmetries. 
All definitions and conditions given 
at the beginning of section \ref{section4}
remain intact modulo the fact that $M=0$ in the case under consideration.
${}$From the Killing  equation 
\eqref{5.7} we deduce  that 
\begin{subequations}
\bea
\z_\a = - \frac{3}{2}
\cD_\a \s| = -\frac{3}{4} \bar \e^\bd [\cD_\a, \bar \cD_\bd ] {\frak L} |
=(b_{\a\bd} + \frac{3}{2} H_{\a\bd} )\bar \e^\bd ~,
\eea
and hence 
\bea
\bar \z^\ad = 
- ( b^{\ad\b} + \frac{3}{2} H^{\ad\b})\e_\b
~.
\eea
\end{subequations}
Plugging this into \eqref{4.11} and  setting $M=0$ gives
\bea
{\mathfrak D}_a \e_\b + \frac{\ri }{2}  (\s_a \tilde{\s}_c \e)_\b H^c  =0  ~.
\label{5.11}
\eea
This equation is invariant under the unbroken U$(1)_R$ gauge group for which $b_a$ is the gauge field
and ${\frak D}_a$ the gauge covariant derivative. 

Eq. \eqref{5.11} coincides with the Killing spinor equation in new minimal supergravity \cite{FS,DFS,KMTZ}.
All results of these papers, which concern supersymmetric backgrounds in new minimal 
supergravity, follow from this equation. The conditions on the background fields implied by 
\eqref{5.11} may be uncovered by studying the corresponding integrability 
conditions, see e.g. \cite{Liu:2012bi}. Alternatively, one may use the superspace formalism 
and analyze implications of the relations 
\begin{subequations}
\bea
0&=& \x^D \cD_D R -2\s R -\frac{1}{4} (\bar \cD^2 -4R ) \bar \s ~, \\
0&=&  \x^D \cD_D G_{\a\ad} +K_\a{}^\d G_{\d\ad} + \bar K_\ad{}^\dd G_{\a\dd} 
 -\hf (\s +\bar \s) G_{\a\ad} +\ri \cD_{\a\ad} (\bar \s- \s) ~,\\
0&=&  \x^D \cD_D W_{\a\b\g} + 3 K^\d{}_{(\a } W_{\b\g)\d} -\frac{3}{2} \s W_{\a\b\g}~,
\eea
\end{subequations}
in conjunction with the following corollary of  \eqref{5.7}:
\bea
0= \x^D \cD_D {\frak H}_{\a\ad} &+&K_\a{}^\d {\frak H}_{\d\ad} + \bar K_\ad{}^\dd {\frak H}_{\a\dd} 
\non \\
& - & \frac{3}{2}(\s+\bar \s) {\frak H}_{\a \ad} +\frac{3}{2} (\cD_\a \s) \bar \cD_\ad {\frak L}
 - \frac{3}{2} (\bar \cD_\ad \bar \s)  \cD_\a {\frak L} ~,
\eea
where we have denoted
\bea
{\frak H}_{\a\ad} :=-\hf( [\cD_\a, \bar \cD_\ad] -2 G_{\a\ad}) {\frak L} ~.
\eea
We will not pursue such an analysis here. 

\subsection{Non-minimal supergravity}
The compensators in  non-minimal supergravity are a complex covariantly linear scalar $\S$
constrained by 
\bea
(\bar \cD^2 -4R )\S =0 ~,
\label{5.12}
\eea
and its complex conjugate. 
The super-Weyl transformation of $\S$ is not determined uniquely by the constraint, 
\bea
\d_\s \S= \Big( \frac{3n-1}{3n+1} \s - \bar \s\Big) \S~,
\label{5.13}
\eea
where $n$ is a real parameter such that $n\neq -1/3, 0$.  
Thus the Killing equation corresponding to non-minimal supergravity is  
\bea
\bar \s -   \frac{3n-1}{3n+1} \s = \x^B\cD_B \ln \S~.
\label{5.18}
\eea

As  has recently  been shown  \cite{BKdual}, 
the only way to construct non-minimal supergravity with 
a cosmological term\footnote{Non-minimal anti-de Sitter supergravity was argued in \cite{GGRS}
not to exist. This is indeed so if one deals with the standard constraint \eqref{5.12}.} 
is to fix $n=-1$ and consider a compensator $\G$ obeying the deformed linearity constraint
\bea
-\frac{1}{4} (\bar \cD^2 - 4 R) \Gamma &= \mu = {\rm const}~.
\label{5.15}
\eea
The super-Weyl transformation of $\G$ is
\bea
\d_\s \G= \big( 2 \s - \bar \s\big) \G~.
\eea
It is obtained from \eqref{5.13} by replacing $\S \to \G$ and setting $n=-1$. 
One may check that the left-hand side of \eqref{5.15} is super-Weyl invariant.
The Killing equation for this supergravity formulation is obtained from \eqref{5.18}
by choosing $n=-1$ and replacing $\S \to \G$.  Anti-de Sitter superspace is a maximally symmetric 
solution of this theory \cite{BKdual}.

\section{Conclusion and outlook}

In this paper we have re-derived some of the key results of \cite{FS,DFS,KMTZ}
from the more general superspace setting developed in \cite{BK}.
The superspace approach of \cite{BK} is more general simply because it takes care of all the isometries
of a given curved background, and not just the rigid supersymmetry transformations 
as in  \cite{FS,DFS,KMTZ}. If one is interested in generating all possible supersymmetric backgrounds
in 4D $\cN=1$ off-shell supergravity, the results of 
\cite{FS,Jia:2011hw,Samtleben:2012gy,Klare:2012gn,DFS,KMTZ,Liu:2012bi,Dumitrescu:2012at,Kehagias:2012fh} appear to be  exhaustive. However, if the goal is to engineer
off-shell rigid supersymmetric theories on a given curved spacetime, 
or to carry out supergraph calculations in such theories, 
the superspace symmetry
formalism of \cite{BK} (and its extensions) is most powerful. In this respect, it is worth mentioning the explicit 
construction of the most general 4D $\cN=2$ supersymmetric sigma models in anti-de Sitter space
\cite{BKLT-M}.

Old minimal supergravity with the super-Weyl invariance can be thought of 
as $\cN=1$ conformal supergravity. From this point of view,  the super-Weyl invariant
approach to all known off-shell $\cN=1$ supergravity theories, 
which we sketched in section \ref{section5}, is simply a version of the general
principle that Poincar\'e (super)gravity can be realized as conformal (super)gravity coupled 
to certain compensators \cite{csg_poincare} (see also \cite{deWR}).
This principle is universal, for it also applies to extended supergravities in four dimensions
and, more generally, to supergravity theories in diverse dimensions. 
Therefore, our approach to the symmetries of curved 4D $\cN=1$ superspace backgrounds
can readily be extended to supergravity theories in diverse dimensions. 
Suitable superspace formulations were constructed in \cite{KLRT-M} for 4D $\cN=2$ 
supergravity, \cite{KT-M_5Dconf} for 5D $\cN=1$ supergravity, \cite{KLT-M11} for 
3D $\cN$-extended supergravity, \cite{Linch:2012zh} for 6D $\cN=(1,0)$ supergravity. 
It is an interesting open problem to  study supersymmetric backgrounds supported by these supergravity theories using the superspace techniques.  
 
Regarding the $\cN=1$ case in four dimensions studied in the present paper, 
there exist alternative superspace formulations for 
conformal supergravity developed by Howe \cite{Howe} (see 
 \cite{GGRS} for a review) and Butter \cite{ButterN=1}, which are characterized by larger structure groups 
 than the Lorentz group
 ${\rm SL}(2,{\mathbb C})$
 characteristic of the Wes--Zumino superspace geometry.\footnote{The formulation 
 for $\cN=1$ conformal supergravity given in 
 \cite{ButterN=1} may be viewed as a master one in the sense that all other formulations
 can be obtained form it by partial gauge fixings,  see \cite{BKdual} for a review.}
 It would be interesting to make use of these formulations to study supersymmetric backgrounds. 
 In particular, since the structure group in Howe's  formulation is  
 ${\rm SL}(2,{\mathbb C}) \times {\rm U}(1)_R$, this approach is most suitable to describe 
 new minimal supergravity. 

While this paper was in the process of writing up, 
there appeared a preprint \cite{Closset:2012ru} devoted to the construction of supersymmetric 
backgrounds associated with one of the off-shell formulations 
for 3D $\cN=2$ supergravity  developed in \cite{KT-M11,KLT-M11}.
The analysis in \cite{Closset:2012ru} is purely component. 
A superspace construction 
may be developed along the lines described in the present paper. 
In fact, the supersymmetric  backgrounds allowing four supercharges were constructed  
in \cite{KT-M11}
much earlier  than \cite{Closset:2012ru} using purely superspace tools.
\\

\noindent
{\bf Acknowledgements:}\\
The author is grateful to Daniel Butter for a useful discussion and for pointing out an error 
in the first draft of this paper. It is also a pleasure to thank Ioseph Buchbinder and Ian McArthur 
for reading the manuscript. 
This work was supported in part by the Australian Research Council, project
No. DP1096372.

\begin{footnotesize}

\end{footnotesize}

\end{document}